# Vcache: Caching Dynamic Documents

Vipul Goyal[1]
Department of Computer Science &
Engg
Institute of Technology, Banaras Hindu
University
Varanasi-221005, India
vipulg@cpan.org

Sugata Sanyal
School of Technology & Computer
Science
Tata Institute of Fundamental Research

Mumbai-400005, India
sanyal@tifr.res.in

Dharma P. Agrawal
Center for Distributed and Mobile
Computing, ECECS
University of Cincinnati

Cincinnati, OH 45221-0030, USA
dpa@ececs.uc.edu

**Abstract** − **The traditional web caching is currently limited to static documents only. A page generated on the fly from a server side script may have different contents on different accesses and hence cannot be cached.**
 **A number of proposals for attacking the problem have emerged based on the observation that different instances of a dynamic document are usually quite similar in most cases, i.e. they have a lot of common HTML code.**
 **In this paper, we first review these related techniques and show their inadequacy for practical use (see section 2). We then present a general and fully automatic technique called Vcache based on the decomposition of dynamic documents into a hierarchy of templates and bindings. The technique is designed keeping in mind languages like Perl and C etc that generate the documents using low-level print like statements. These languages together, account for the largest number of dynamic documents on the web.**

## 1. INTRODUCTION

The World Wide Web is comprised of static documents and dynamic documents. One central aspect of the development of the WWW during the last decade is the increasing use of dynamic documents [1, 16]. More and more HTML documents are dynamically generated on the fly using a server side script, which are commonly written in Perl, C, Php, Asp and Jsp etc.

The traditional caching fails in case of dynamic documents, since every instance is generated on the fly using a server side script and generally cannot be assumed to be the same as the previous instance. Hence, since a dynamic document instance should be downloaded in its entirety for each request, the network bandwidth consumed and more importantly the response time is substantial.

The motivation for our caching scheme is the observation that the different instances of dynamic document may differ only slightly in content and usually contain a number of sections of common HTML code.

 The central idea of Vcache is to decompose the dynamic document into a hierarchy of templates and bindings. The templates can then be cached at the client side while the bindings are supplied by the server separately for each instance of that dynamic document. This decomposition of dynamic documents is automatic. The decomposition is done by a software running at the server using branch flow statistics technique. We will call this server software as the fragmentor from hereon.

## II. RELATED WORKS

The techniques labelled "dynamic document caching" can be broadly classified into three categories: server based, proxy based and client based. Though a number of server- based and proxy-based techniques are available (e.g. [11, 6, 9, 5, 14]), not many are client based. The technique we propose here is a client-based technique e.g. [7, 12].

Delta encoding [10] is based on the observation that most dynamically constructed documents have many fragments in common with earlier versions. Instead of transferring the complete document, a delta is computed representing the changes compared to some common base. Using a cache proxy, the full document is regenerated near the client. A drawback is—in addition to requiring specialized proxies— that it necessitates protocols for management of past versions. Such intrusions can obviously limit widespread use. Furthermore, it does not help with repetitions within a single document.

In the <bigwig> system [12], dynamic HTML documents are composed of higher-order templates that are plugged together to construct complete documents. A <bigwig> service transmits not the full HTML document but instead a compact JavaScript recipe for a client-side construction of the document based on a static collection of fragments that can be cached by the browser in the usual manner. This technique exploits the template mechanism in the <bigwig> language and hence cannot be extended to other languages without an inbuilt template mechanism like Perl, C and Jsp.

HPP [7] is an HTML extension, which allows an explicit separation between static and dynamic parts of a dynamically generated document. The static parts of a document are collected in a template file while the dynamic parameters are in a separate binding file. The template file can contain simple instructions, akin to embedded scripting languages such as ASP, PHP, or JSP, specifying how to assemble the complete document. In the HPP system, the document construction should be explicitly programmed and each document should be manually divided into templates and bindings. The HPP system will require every dynamic document already present on the web to be reprogrammed in order to benefit from caching. Future documents should also be programmed keeping in mind the HPP caching system and document construction at the client side.

Our solution does not suffer from these drawbacks. The division of the dynamic documents into templates and bindings is fully automatic and applies to the existing

---

documents as well without demanding any changes to them. The programmer need not be aware of the document division/construction or the HTML extension

Vcache requires extra functionality from clients and the server as in HPP. The client functionality can be in the form of either cache proxies or browser plug-ins (for document construction using templates and bindings). The server software should be modified to include the fragmentor. No extension to the HTTP protocol however, is required.

## III. THE CACHING SCHEME

### 3.1 Templates and Bindings
First we define templates and bindings-

**Definition 1 (Template)**
A template is a cacheable regular HTML file having gaps or discontinuities in it. Apart from the regular HTML code, it may contain the following new tags-
1) <gap>
2) <loop> and </loop>

**Definition 2 (Binding)**
A binding is a non-cacheable section of code enclosed between <temp ref="<absolute template url>"> and </temp> tags. The <temp ref="..."> tag specifies the template to which the binding belongs. The enclosed code may contain the following apart from the regular HTML code-
1) <gap> and </gap> tags
2) <loop> and </loop> tags
3) <n> and </n> tags where n is a positive integer not equal to zero.
4) Another binding

These definitions and the meaning of the new tags will be clear from the definition of Plug operator (see next subsection).

### 3.2 Operators required at the client side
As described before, our solution requires extra functionality from the client as in HPP. This can be in the form of cache proxies or java applets or browser plug-ins. The client is required to support a set of operators which are defined below-

**Definition 3 (The GenerateList operator)**
The GenerateList operator takes a binding as a parameter and returns a list of url's of templates. These are the templates, which are referred in that binding and hence will be required for document construction.
This operator is essentially a recursive parser due to the possibility of another binding in a binding (see definition 2).

**Definition 4 (The FetchList operator)**
The FetchList operator accepts the template URL list generated by the GenerateList operator. It then checks the browser cache and in case a subset of these templates is not already present in the cache, it fetches and saves that subset to the cache.

**Definition 5 (The Plug operator)**
The Plug operator accepts a binding and a template as parameters and returns a regular HTML code constructed from them. The Plug operation proceeds according to the following simple rules-

1) Every <gap> tag in the template is replaced by a HTML code section enclosed between <gap> and </gap> tags in the binding. This replacing is sequential.
2) The <loop> and </loop> tags and the enclosed code in the template are replaced by the HTML code generated by plugging the inside code separate with each run of the loop specified in the binding. The nth run of the loop is specified by the code enclosed inside <n> and </n> tags in the binding and all the runs together are enclosed between <loop> and </loop> tags (see Figure 1).
3) If a binding enclose more binding(s) (which may again enclose more binding(s) and so on) within the <temp ref="..."> and </temp> tags, these tags and the binding(s) inside are replaced by the code generated after applying the Plug operator recursively to that binding and the template specified in the <temp ref="..."> tag.

The illustration given (see Figure 1) should make the Plug operator clear.

### 3.3 The actual caching mechanism
The fragmentor produces a set of templates from the dynamic document at the server (see next subsection). These templates are analogous to the static documents and can be cached by the client.

For every access to the dynamic document, a binding is generated at the server and is passed to the client. The binding is specific to that particular access and hence is non-cacheable. The client then fetches the un-cached templates if any (using the GenerateList and FetchList operators). The binding and its template (as in <temp ref="...") are then passed to the Plug operator which recursively generates the document instance, which is then finally shown to the end user.

### 3.4 The fragmentor (fragmentor operator)
The fragmentor as described before is integrated with the server software and is mainly responsible for the automatic division of the dynamic documents into a hierarchy of templates and bindings.

The fragmentor can produce a number of templates simply by parsing the source code of the script (dynamic document). While parsing, the print output of a variable (whose value will be different for different runs of the script) in the template is replaced by a <gap> tag. On encountering a branching decision (if statement etc), a <gap> tag is inserted in the template and separate templates are constructed recursively for every possible branch at that point. This approach is illustrated in Figure 2.

However, even for moderately sized programs, this brute force approach of template generation may produce a large number of templates. The number of templates produced is equal to the number of branch flow possibilities of the program. This places unnecessary burden on the system as some of the templates are rarely used (e.g. error/exception handling branches) and some others may be too small in size.

We therefore take a variant of this approach to optimize the system. The fragmentor (after getting installed on the server) first gathers the branch flow statistics of the dynamic documents on the server by analysing the client requests and execution of server scripts (dynamic documents) to provide the response. After collecting a meaningful statistics (say for n runs of the script), the fragmentor generates one template each for all the dominating branch sequences/subsequences instead of generating separate template for every possible branch. Hence effectively, templates for rarely taken branches are not generated at all, while the templates for a popular branching sequence get merged into a single template (see Figure 3).

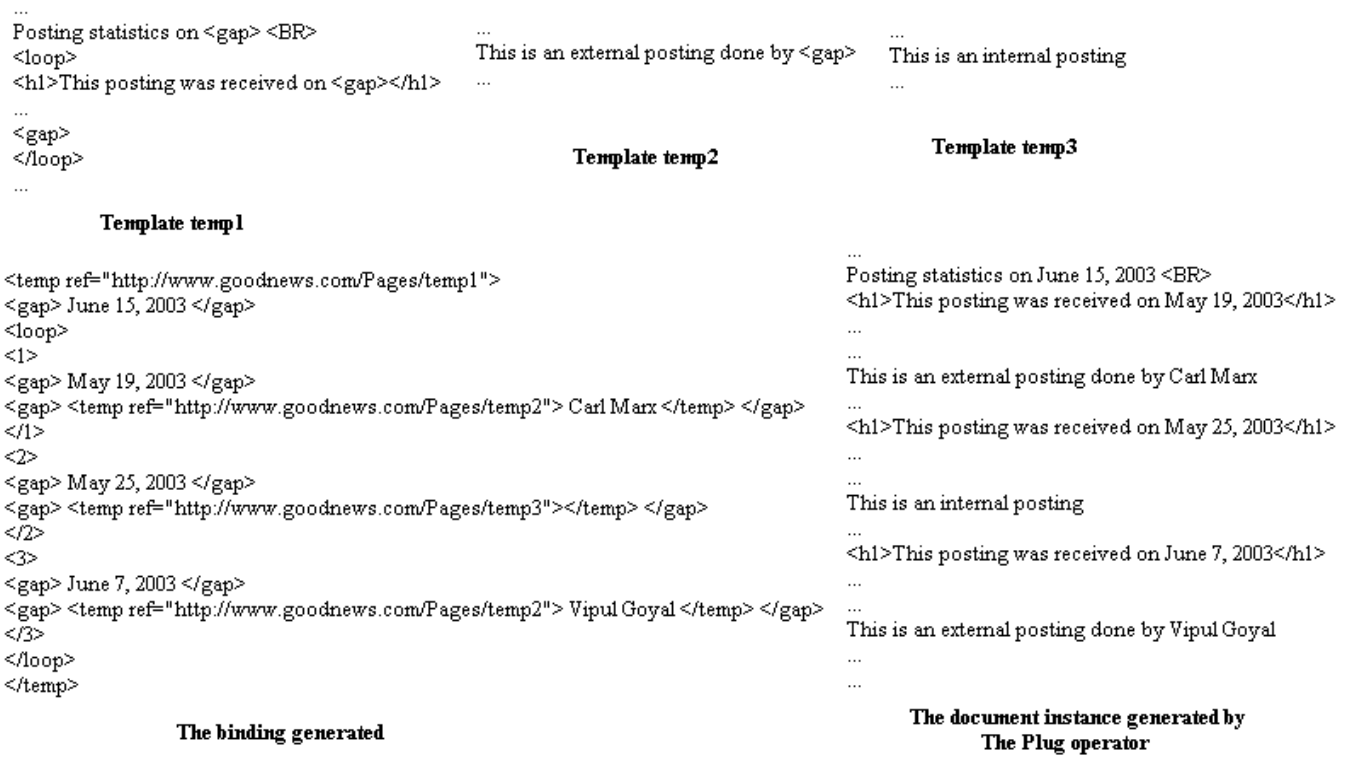

Figure 1

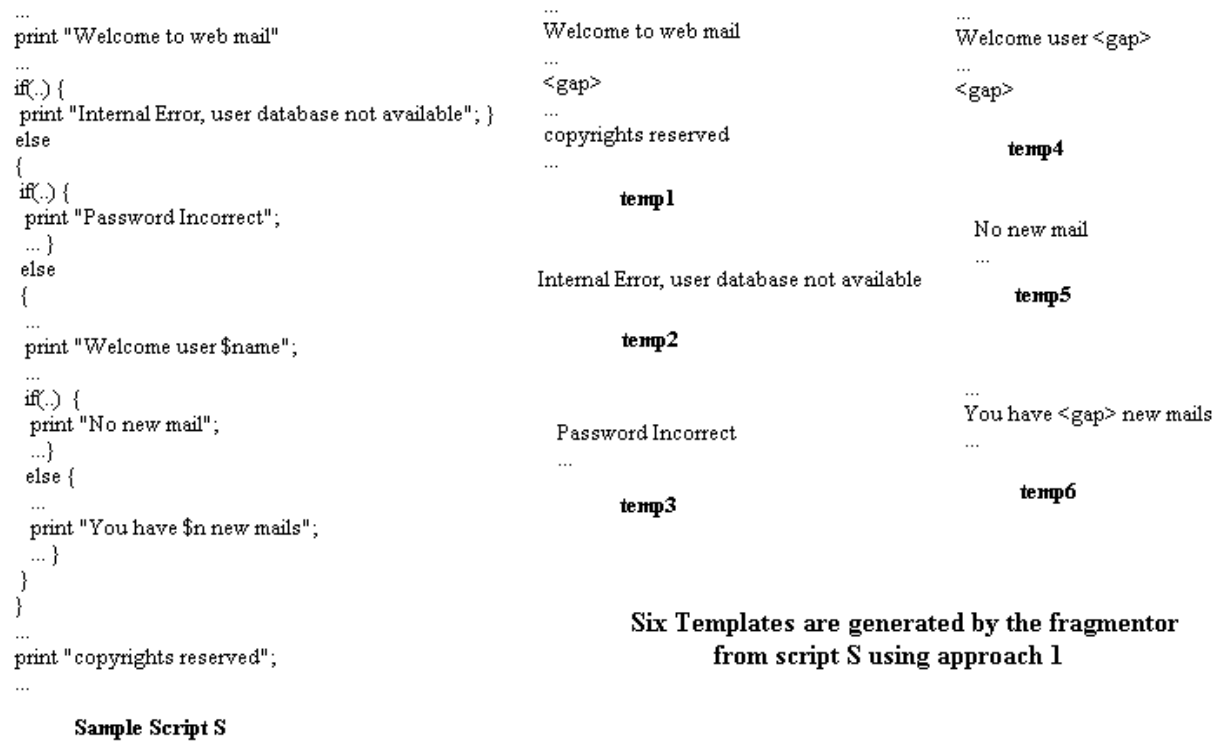

Figure 2

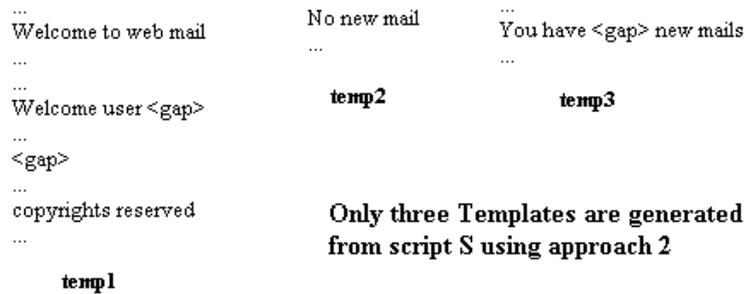

...
Welcome to web mail
...
Welcome user <gap>
...
<gap>
...
copyrights reserved
...

**temp1**

No new mail
...

**temp2**

...
You have <gap> new mails
...

**temp3**

**Only three Templates are generated
from script S using approach 2**

**Figure 3**

Also, the fragmentor does not generate the templates whose size is below a lower limit (say 50 bytes) or duplicate templates (inter-document optimization).

We are currently working to make a detailed design for the fragmentor using various optimizations and to implement it for Perl language. The implementation is yet to be finished; hence the formal performance metrics for our technique are currently unavailable. However, it can be expected that with a nicely written and optimizing fragmentor, the performance should be at par with the techniques like <bigwig> and HPP while being general and automatic at the same time.

## IV. CONCLUSION

We have proposed a caching technique, which we call Vcache, to extend the existing client side caching mechanism to cover dynamic documents. Our approach requires changes to the server and the client software. No change to the HTTP protocol is required. Our approach is general and fully automatic and can be used for dynamic documents designed in languages like Perl, C, Jsp etc. The template hierarchy is generated using the real run time data of the server scripts. With our technique, the programmer need not be aware of the caching issues since the decomposition of the dynamic documents into templates and bindings is automatic.